\documentclass[pra, showpacs, twocolumn, floatfix,superscriptaddress]{revtex4}
\usepackage{graphicx}
\usepackage{epstopdf}
\usepackage{amsmath, amsfonts, amssymb, bm}

\def\mpik{Max-Planck-Institut f\"ur Kernphysik, Saupfercheckweg 1, D-69117
Heidelberg, Germany}

\def\ifa{Institute of Applied Physics, Academy of Sciences of Moldova,
Academiei str. 5, MD-2028 Chi\c{s}in\u{a}u, Moldova}
\begin{document}
%%%%%%%%%%%%%%%%%%%%%%%%%%%%%%%%%%%%%%%%%%%%%%%%%%%%%%%%%%%%%%%%%
\title{Quantum dynamics of a two-level emitter with modulated transition frequency}
%%%%%%%%%%%%%%%%%%%%%%%%%%%%%%%%%%%%%%%%%%%%%%%%%%%%%%%%%%%%%%%%%
\author{Mihai \surname{Macovei}}
\email{macovei@phys.asm.md}
\affiliation{\mpik}
\affiliation{\ifa}

\author{Christoph H. \surname{Keitel}}
\email{keitel@mpi-hd.mpg.de}
\affiliation{\mpik}

\date{\today}
%%%%%%%%%%%%%%%%%%%%%%%%%%%%%%%%%%%%%%%%%%%%%%%%%%%%%%%%%%%%%%%%%
\begin{abstract}
The resonant quantum dynamics of an excited two-level emitter is 
investigated via classical modulation of its transition frequency while 
simultaneously the radiator interacts with a broadband electromagnetic 
field reservoir. The 
frequency of modulation is selected to be of the order of the bare-state 
spontaneous decay rate. In this way, one can induce quantum interference 
effects and, consequently, quantum coherences among multiple decaying 
transition pathways. Depending on the modulation depth and its absolute 
phase, both the spontaneous emission and the frequency shift may  be
conveniently modified and controlled. 
\end{abstract}
%%%%%%%%%%%%%%%%%%%%%%%%%%%%%%%%%%%%%%%%%%%%%%%%%%%%%%%%%%%%%%%%%%%
\maketitle
%%%%%%%%%%%%%%%%%%%%%%%%%%%%%%%%%%%%%%%%%%%%%%%%%%%%%%%%%%%%%%%%%%%

%%%%%%%%%%%%%%%%%%%%%%%%%%%%%%%%%%%%%%%%%%%%%%%%%%%%%%%%%%%%%%%%%%%
\section{Introduction}
%%%%%%%%%%%%%%%%%%%%%%%%%%%%%%%%%%%%%%%%%%%%%%%%%%%%%%%%%%%%%%%%%%%
Spontaneous emission is a well established fundamental phenomenon 
\cite{gsa_book,al_eb,ficek,kmek}. It occurs due to interaction of 
excited emitters with the vacuum modes of the environmental 
electromagnetic field reservoir. 
Useful applications of spontaneous radiation control may arise, for instance, in higher frequency 
coherent light generation \cite{x_ray,xx_ray} or spontaneous parametric down conversion processes \cite{w_mil}.
On the other side, spontaneous emission often 
plays a negative role in quantum processing of information \cite{ni_ch}. 
Therefore, it is not surprising that a significant amount of work is carried out
regarding its control. Particularly, earlier approaches to influence the 
spontaneous emission were by using optical cavities \cite{purcel,klepp,haroche}. 
A modern and more advanced version of those ideas consists in using 
photonic crystals environments where photon forbidden bands occur 
leading to spontaneous emission inhibition or localization 
\cite{phkr1,phkr2,phkr3}. Infrequent application to a two-level atom 
of microwave pulses \cite{shap} or sequence of pulses \cite{lloyd}, or 
rather intense low-frequency coherent fields \cite{ek} (see also \cite{ekc}) 
lead to spontaneous emission control as well. 
Quenching of spontaneous emission occurs as well via involving quantum 
interference effects between various decaying pathways which are 
dependent on mutual orientation of corresponding transition dipoles 
\cite{kmek,qin,joerg}. Furthermore, the Lamb shift of laser-dressed 
atomic states and quantum interferences due to energy shifts and their 
effect on spontaneous emission were investigated too, in Refs.~\cite{ulri,zb}. 
One can also control the spontaneous emission by periodically shifting 
the atomic transition frequency from the atom-cavity resonance 
\cite{gsa,jano} or via coupling a single state to a continuum of many states
\cite{kur,asw}. Remarkably, periodically perturbed atomic transitions lead 
to a number of fascinating effects such as induced transparency or extreme 
ultra-short pulses, respectively \cite{koch}.
%%%%%%%%%%%%%%%%%%%%%%%%%%%%%%%%%%%%%%%%%%%%%%%%%%%%%%%%%%%%%%%%%%%%%%%%%%%%%%
\begin{figure}[b]
\includegraphics[width=7.5cm]{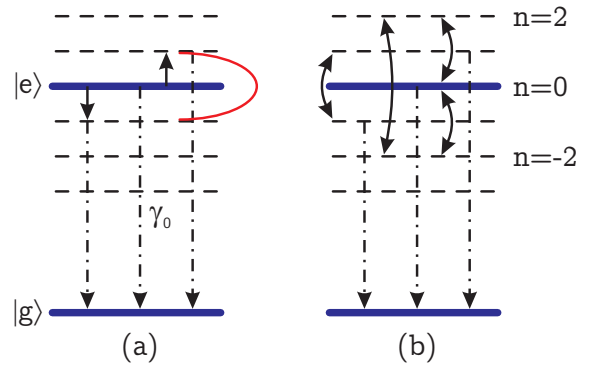}
\caption{\label{fig1} (color online) Schematic diagram of a two-level emitter with modulated 
transition frequency: (a) without and (b) with showing the involved quantum coherences 
among specific transition pathways. Due to modulation, multiple induced decaying channels, 
$n= 0, \pm 1,  \pm 2, \cdots$, interfere such that spontaneous emission is slowed down. 
$\gamma_{0}$ is the single-qubit spontaneous decay rate in the absence of modulation.}
\end{figure}
%%%%%%%%%%%%%%%%%%%%%%%%%%%%%%%%%%%%%%%%%%%%%%%%%%%%%%%%%%%%%%%%%%%%%%%%%%%%%%

Here, we demonstrate suppression of spontaneous decay of a two-level system (qubit) that is embedded in 
a broadband electromagnetic field reservoir and is subjected to an intense, time-dependent, frequency modulation 
driving force. The suppression is a direct consequence of quantum interference effects induced by the modulation. 
A frequency shift to the transition frequency is induced as well. Furthermore, the absolute phase of the modulation can be a 
convenient tool to control these processes. Coherent modulation of the transition 
frequency leads to appearance of new decay channels that may interfere 
destructively contributing to spontaneous emission inhibition (see Fig.~\ref{fig1}). 
This occurs when the frequency of modulation is of the order of the bare-state 
qubit's decay rate or less. The spontaneous emission is described by an exponential 
decaying law with a time-dependent decay rate and exhibiting plateaus with a very 
slow decoherence rate. The quantum decoherence due to spontaneous emission can 
be further minimized via stronger frequency modulation depths. Moreover, the induced 
time-dependent frequency shift depends on external control parameters, such as the applied 
intensity and the external field amplitude absolute phase, and can be influenced accordingly. It 
vanishes, however, at resonance and in the absence of quantum interference effects 
due to frequency modulation processes. In a free-space setup, the spontaneous emission 
inhibition is less probable via transition frequency modulation of an excited two-level emitter.
This deviation to \cite{ek} arises because our treatment is classical and especially limited to 
moderately intense modulating fields.

Our system can be implemented, for example, via off-resonant laser driving of a two-level emitter 
\cite{ek,gsa,jano,kur,asw,koch,exp}. One can apply a laser field with a high non-resonant frequency 
$\omega^{'}$ and possessing a periodically modulated amplitude of the field strength 
$\epsilon(t)=\epsilon_{0}\cos{(\omega t + \phi)}\cos{(\omega^{'} t)}$ with $\omega \ll \omega_{0} \ll \omega^{'}$ 
to a two-level atom of frequency $\omega_{0}$. Then a modulated shift $\Delta \omega_{0}$ of the 
transition frequency is achieved via the quadratic Stark effect, i.e., 
$\Delta \omega_{0}$=$b\cos^{2}{(\omega t + \phi)}$, where $b$ is the modulation amplitude.
Additional systems can be molecules or quantum dots, even those possessing permanent dipoles 
\cite{kov,kibis,okm,shel,pasp_gr,kch}. When pumped with an intense low-frequency 
coherent field, the amplitude of the frequency modulation will be proportional to the 
magnitude of the permanent dipole multiplied by the external field strength. An 
alternative scheme can be as well a two-level quantum dot embedded in a broadband 
microcavity and interacting with a surface acoustic wave coherently modulating its 
transition frequency \cite{ac_w}. Superconducting qubits with periodically perturbed 
transition frequencies and weakly coupled with a quantum LC circuit or a 
nanomechanical resonator are suitable candidates, too \cite{shev}. 

The article is organized as follows. In Sec. II we describe
the analytical approach and the system of interest, while in
Sec. III we analyse the obtained results. The summary given in Sec. IV is 
followed by two Appendixes.
%%%%%%%%%%%%%%%%%%%%%%%%%%%%%%%%%%%%%%%%%%%%%%%%%%%%%%%%%%%%%%%%%%%%%%%%%%
\section{Approach}
%%%%%%%%%%%%%%%%%%%%%%%%%%%%%%%%%%%%%%%%%%%%%%%%%%%%%%%%%%%%%%%%%%%%%%%%%%
The Hamiltonian, $H=H_{0} + H_{I}$, describing the system of interest can 
be represented via (see \cite{gsa,jano} or for a detailed derivation for 
the analogous case with usual vacuum Appendix A):
%%%%%%%%%%%%%%%%%%%%%%%%%%%%%%%%%%%%%%%%%%%%%%%%%%%%%%%%%%%%%%%%%%%%%%%%%%
\begin{eqnarray}
H_{0} &=& \sum_{k}\hbar \omega_{k}a^{\dagger}_{k}a_{k} 
+ \hbar\bigl(\omega_{0} + b\cos{(\omega t + \phi)}\bigr)S_{z}, \nonumber \\
H_{I} &=& i\sum_{k}(\vec g_{k}\cdot \vec d)\bigl (a^{\dagger}_{k}S^{-} 
- a_{k}S^{+}\bigr), \label{Hh}
\end{eqnarray}
%%%%%%%%%%%%%%%%%%%%%%%%%%%%%%%%%%%%%%%%%%%%%%%%%%%%%%%%%%%%%%%%%%%%%%%%%%
where the first term, $H_{0}$, characterizes the free Hamiltonian of the electromagnetic 
field (EMF) as well as of the qubit subsystem with modulated transition frequency whereas 
the second one, i.e. $H_{I}$, accounts for the interaction of the two-level qubit with 
the vacuum modes of the environmental electromagnetic field reservoir. 
Here, $S^{+}=|e\rangle\langle g|$, $S^{-}=[S^{+}]^{\dagger}$ and 
$S_{z}=(|e\rangle\langle e| - |g \rangle\langle g|)/2$ are 
the well-known quasispin operators obeying the commutation relations: 
$[S^{+},S^{-}]=2S_{z}$ and $[S_{z},S^{\pm}]=\pm S^{\pm}$. The creation 
$a^{\dagger}_{k}$ and annihilation $a_{k}$ electromagnetic field operators 
satisfy the commutation relations: $[a_{k},a^{\dagger}_{k'}]=\delta_{kk'}$ 
and $[a_{k},a_{k'}]=[a^{\dagger}_{k},a^{\dagger}_{k'}]=0$. Further, 
$\omega_{0}$ is the qubit's transition frequency 
$|e \rangle \leftrightarrow |g \rangle$ (see Fig.~\ref{fig1}) in the absence 
of classical modulation, while $b$ is the modulation amplitude with frequency $\omega$ 
and phase $\phi$. The two-level emitter possessing the transition dipole moment 
$d$ couples with the vacuum modes via the coupling constant $g_{k}$. 
In the following, we perform a unitary transformation:
%%%%%%%%%%%%%%%%%%%%%%%%%%%%%%%%%%%%%%%%%%%%%%%%%%%%%%%%%%%%%%%%%%%%%%%%%%
\begin{eqnarray}
U = \exp\{\frac{i}{\hbar}\int^{t}_{0}d\tau \bar H_{0}(\tau)\}, \label{UT}
\end{eqnarray}
%%%%%%%%%%%%%%%%%%%%%%%%%%%%%%%%%%%%%%%%%%%%%%%%%%%%%%%%%%%%%%%%%%%%%%%%%%
with $\bar H_{0}(\tau) = \sum_{k}\hbar \omega_{0}a^{\dagger}_{k}a_{k} 
+ \hbar\bigl(\omega_{0} + b\cos{(\omega \tau + \phi)}\bigr)S_{z}$ and arrive
at the Hamiltonian:
%%%%%%%%%%%%%%%%%%%%%%%%%%%%%%%%%%%%%%%%%%%%%%%%%%%%%%%%%%%%%%%%%%%%%%%%%%
\begin{eqnarray}
\tilde H &=& \sum_{k}\hbar(\omega_{k}-\omega_{0})a^{\dagger}_{k}a_{k} + 
i\sum_{k}\sum^{\infty}_{m = -\infty}(\vec g_{k}\cdot \vec d)J_{m}(\chi) \nonumber \\
&\times&\bigl (a^{\dagger}_{k}S^{-} e^{-im(\omega t+\phi)} -  
a_{k}S^{+}e^{im(\omega t+\phi)} \bigr),  \label{Heff}
\end{eqnarray}
%%%%%%%%%%%%%%%%%%%%%%%%%%%%%%%%%%%%%%%%%%%%%%%%%%%%%%%%%%%%%%%%%%%%%%%%%%
where $\chi=b/\omega$ while $J_{m}(\chi)$ is the corresponding ordinary Bessel function. 
Here, we used the expansion via the $m$th-order Bessel function of the first kind, i.e., 
$\exp\{\pm i\chi \sin{(\omega t + \phi)}\} = 
\sum^{\infty}_{m=-\infty}J_{m}(\chi)\exp{\bigl(\pm im(\omega t + \phi)\bigr)}$ 
as well as the notation: $S^{\pm}e^{\mp i\chi\sin\phi} \equiv \tilde S^{\pm}$, 
and dropped the tilde afterwards. 

In the weak qubit-environment coupling limit, one can obtain the master equation 
describing the quantum dynamics of any atomic operator $Q$. For this, we use the 
standard elimination procedure of the electromagnetic field operators from the 
Heisenberg equation: 
%%%%%%%%%%%%%%%%%%%%%%%%%%%%%%%%%%%%%%%%%%%%%%%%%%%%%%%%%%%%%%%%%%%%%%%%%%
\begin{eqnarray}
\frac{d}{dt}\langle Q\rangle = \frac{i}{\hbar}\langle [\tilde H,Q]\rangle, \label{eqQ}
\end{eqnarray}
%%%%%%%%%%%%%%%%%%%%%%%%%%%%%%%%%%%%%%%%%%%%%%%%%%%%%%%%%%%%%%%%%%%%%%%%%%
where the notation $\langle \cdots \rangle$ indicates averaging over the initial
state of both the qubit and the surrounding electromagnetic field bath
\cite{gsa_book,al_eb,ficek,kmek}. 
As an environmental electromagnetic field reservoir, we consider a broadband optical 
cavity possessing the frequency $\omega_{c}$, qubit-cavity coupling being $g$, and a 
cavity leaking constant denoted by $\kappa$ (the free-space situation is described in Appendix A). 
Thus, the Heisenberg equations for the field operators are:
%%%%%%%%%%%%%%%%%%%%%%%%%%%%%%%%%%%%%%%%%%%%%%%%%%%%%%%%%%%%%%%%%%%%%%%%%%
\begin{eqnarray}
\frac{d}{dt}a^{\dagger}(t) = (i\delta_{c}-\kappa)a^{\dagger} + \sum^{\infty}_{n=-\infty}gJ_{n}(\chi)S^{+}(t)e^{in(\omega t + \phi)}, \label{heq}
\end{eqnarray}
%%%%%%%%%%%%%%%%%%%%%%%%%%%%%%%%%%%%%%%%%%%%%%%%%%%%%%%%%%%%%%%%%%%%%%%%%%
with $a(t)=[a^{\dagger}(t)]^{\dagger}$ and $\delta_{c}=\omega_{c}-\omega_{0}$. Its formal solution in the weak-coupling limit is 
$a^{\dagger}(t)=a^{\dagger}_{v}(t)$+$a^{\dagger}_{s}(t)$, where $a^{\dagger}_{v}(t)=a^{\dagger}(0)e^{-(\kappa - i\delta_{c})t}$ while 
%%%%%%%%%%%%%%%%%%%%%%%%%%%%%%%%%%%%%%%%%%%%%%%%%%%%%%%%%%%%%%%%%%%%%%%%%%
\begin{eqnarray}
a^{\dagger}_{s}(t)&=&\sum^{\infty}_{n=-\infty}gJ_{n}(\chi)\int^{t}_{0}dt^{'}e^{-(\kappa - i\delta_{c})(t-t^{'})}\nonumber \\
&\times& S^{+}(t^{'})e^{in(\omega t^{'}+\phi)}. \label{as}
\end{eqnarray}
%%%%%%%%%%%%%%%%%%%%%%%%%%%%%%%%%%%%%%%%%%%%%%%%%%%%%%%%%%%%%%%%%%%%%%%%%%
In the Markov approximation we have $S^{+}(t^{'}) \approx S^{+}(t)$. Then the integral
%%%%%%%%%%%%%%%%%%%%%%%%%%%%%%%%%%%%%%%%%%%%%%%%%%%%%%%%%%%%%%%%%%%%%%%%%%
\begin{eqnarray*}
\int^{t}_{0}dt^{'}e^{(\kappa - i\delta_{c})t^{'}}e^{in\omega t^{'}} 
= \frac{e^{(\kappa - i(\delta_{c}-n\omega))t}-1}{\kappa + i(n\omega - \delta_{c})}.
\end{eqnarray*}
%%%%%%%%%%%%%%%%%%%%%%%%%%%%%%%%%%%%%%%%%%%%%%%%%%%%%%%%%%%%%%%%%%%%%%%%%%
Inserting this expression in Eq.~(\ref{as}) and keeping only the slower contributions, that is, we are interested in 
frequency modulation regimes slower than the cavity decay rate, i.e. $\omega \ll \kappa$,  one arrives at:
%%%%%%%%%%%%%%%%%%%%%%%%%%%%%%%%%%%%%%%%%%%%%%%%%%%%%%%%%%%%%%%%%%%%%%%%%%
\begin{eqnarray}
a^{\dagger}(t) &=& a^{\dagger}(0)e^{-(\kappa - i\delta_{c})t} + \sum^{\infty}_{n=-\infty}\frac{gJ_{n}(\chi)}{\kappa + i(n\omega - \delta_{c})}\nonumber \\
&\times& S^{+}(t)e^{in(\omega t + \phi)}. \label{af}
\end{eqnarray}
%%%%%%%%%%%%%%%%%%%%%%%%%%%%%%%%%%%%%%%%%%%%%%%%%%%%%%%%%%%%%%%%%%%%%%%%%%
Then, one can write down the master equation for an arbitrary mean-value of a qubit operator $Q$ that can be obtained 
after introducing Eq.~(\ref{Heff}) in the corresponding Heisenberg equation, i.e. Eq.~(\ref{eqQ}):
%%%%%%%%%%%%%%%%%%%%%%%%%%%%%%%%%%%%%%%%%%%%%%%%%%%%%%%%%%%%%%%%%%%%%%%%%%
\begin{eqnarray}
\langle \dot Q\rangle &=& - \sum^{\infty}_{m=-\infty}gJ_{m}(\chi)\{\langle a^{\dagger}[S^{-},Q]\rangle e^{-im(\omega t + \phi)} \nonumber \\
&+& \langle [Q,S^{+}]a\rangle e^{im(\omega t + \phi)}\}, \label{MQ1}
\end{eqnarray}
%%%%%%%%%%%%%%%%%%%%%%%%%%%%%%%%%%%%%%%%%%%%%%%%%%%%%%%%%%%%%%%%%%%%%%%%%%
where an overdot denotes differentiation with respect to time. Introducing Eq.~(\ref{af}) in the master equation 
(\ref{MQ1}) and taking into account that $\langle a^{\dagger}(0)\cdots \rangle=0$ and $\langle \cdots a(0)\rangle =0$ one obtains:
%%%%%%%%%%%%%%%%%%%%%%%%%%%%%%%%%%%%%%%%%%%%%%%%%%%%%%%%%%%%%%%%%%%%%%%%%%
\begin{eqnarray}
\langle \dot Q\rangle &=& i\Omega(t)\langle [S_{z},Q]\rangle - \gamma(t)\{\langle S^{+}[S^{-},Q]\rangle  \nonumber \\
&+& \langle [Q,S^{+}]S^{-}\rangle \}. \label{MQ}
\end{eqnarray}
%%%%%%%%%%%%%%%%%%%%%%%%%%%%%%%%%%%%%%%%%%%%%%%%%%%%%%%%%%%%%%%%%%%%%%%%%%
The operator form of Eq.~(\ref{MQ}) looks standard, i.e. of Lindblad form \cite{ficek,lind}, 
with, however, time-dependent coefficients, namely:
%%%%%%%%%%%%%%%%%%%%%%%%%%%%%%%%%%%%%%%%%%%%%%%%%%%%%%%%%%%%%%%%%%%%%%%%%%
\begin{eqnarray}
\Omega(t)&=& \sum^{\infty}_{\{m,n \} =-\infty}\bar \delta_{n}J_{m}(\chi)
J_{n}(\chi)\cos{[(n-m)(\omega t + \phi)]}, \nonumber \\
\gamma(t) &=& \sum^{\infty}_{\{m,n \} =-\infty}\bar \gamma_{n}J_{m}(\chi)
J_{n}(\chi)\cos{[(n-m)(\omega t + \phi)]}, \nonumber \\ \label{kf}
\end{eqnarray}
%%%%%%%%%%%%%%%%%%%%%%%%%%%%%%%%%%%%%%%%%%%%%%%%%%%%%%%%%%%%%%%%%%%%%%%%%%%%%%
with
%%%%%%%%%%%%%%%%%%%%%%%%%%%%%%%%%%%%%%%%%%%%%%%%%%%%%%%%%%%%%%%%%%%%%%%%%%
\begin{eqnarray}
\bar \delta_{n}=\frac{(n\omega - \delta_{c})g^{2}}{\kappa^{2}+(n\omega - \delta_{c})^{2}}, 
~{\rm and}~\bar \gamma_{n}=\frac{\gamma_{0}\kappa^{2}}{\kappa^{2}+(n\omega - \delta_{c})^{2}}. 
\label{kdg}
\end{eqnarray}
%%%%%%%%%%%%%%%%%%%%%%%%%%%%%%%%%%%%%%%%%%%%%%%%%%%%%%%%%%%%%%%%%%%%%%%%%%%%%%
Here, $\Omega(t)$ and $\gamma(t)$ describe the time-dependent frequency shift 
and spontaneous decay process, respectively, while $\gamma_{0} = g^{2}/\kappa$ is 
the near-resonance single-qubit spontaneous decay rate without frequency 
modulation, i.e. when $\chi=0$. In the numerical simulations we truncate the summation 
range $(-\infty,\infty)$ to $(-\bar n,\bar n)$.
This is justified, as $\bar \gamma_{n} \sim 1/[\kappa^{2} + (n\omega)^{2}]$ for near 
qubit-cavity resonance. Concretely, $\bar n$ is chosen such that the results 
converge, i.e. remain unchanged if one further increases $\bar n$. Note that this is not 
the case for vacuum free-space setups (see Appendix A). Furthermore, in order to avoid 
unphysical results \cite{bar_sten}, $\bar n$ should be the same for both indices and, also, 
Eq.~(\ref{MQ}) should be independent on exchange of indices, i.e. $m \leftrightarrow n$. 

The population quantum dynamics of an initially excited two-state radiator can 
be easily obtained from Eq.~(\ref{MQ}), namely:
%%%%%%%%%%%%%%%%%%%%%%%%%%%%%%%%%%%%%%%%%%%%%%%%%%%%%%%%%%%%%%%%%%%%%%%%%%%%%%
\begin{eqnarray}
\langle S_{z}(t)\rangle &=& \exp{[-2\Gamma(t)]} - 1/2,
\label{sz}
\end{eqnarray}
%%%%%%%%%%%%%%%%%%%%%%%%%%%%%%%%%%%%%%%%%%%%%%%%%%%%%%%%%%%%%%%%%%%%%%%%%%%%%%
with a generalized spontaneous decay rate given by:
\begin{eqnarray*}
\Gamma(t)=\int^{t}_{0}\gamma(\tau)d\tau.
\end{eqnarray*}
One can see here that the qubit inversion obeys a modified exponential decay
law with a time-dependent decay rate. In the absence of frequency modulation,
i.e. $\chi=0$, one recovers the standard exponential law near qubit-cavity 
resonance \cite{kmek}: 
%%%%%%%%%%%%%%%%%%%%%%%%%%%%%%%%%%%%%%%%%%%%%%%%%%%%%%%%%%%%%%%%%%%%%%%%%%%%%%
\begin{eqnarray}
\langle S_{z}(t)\rangle =\exp{[-2\gamma_{0} t]} - 1/2. 
\label{sz0}
\end{eqnarray}
%%%%%%%%%%%%%%%%%%%%%%%%%%%%%%%%%%%%%%%%%%%%%%%%%%%%%%%%%%%%%%%%%%%%%%%%%%%%%%
Thus, the periodical modulation of the qubit's transition frequency modifies the spontaneous 
decay. 

In the following Section, we shall describe the quantum dynamics of an excited two-level 
emitter with modulated transition frequency.

%%%%%%%%%%%%%%%%%%%%%%%%%%%%%%%%%%%%%%%%%%%%%%%%%%%%%%%%%%%%%%%%%%%%%%%%%%%%%%
\section{Results and Discussion}
%%%%%%%%%%%%%%%%%%%%%%%%%%%%%%%%%%%%%%%%%%%%%%%%%%%%%%%%%%%%%%%%%%%%%%%%%%%%%%
We proceed to investigate the qubit's dynamics based on Eqs.~(\ref{af}-\ref{sz}). 
One can observe from Eq.~(\ref{af}) that the atomic dipole may oscillate at frequencies  
$\omega_{n} = \omega_{0} + n\omega$, where $n$ is an arbitrary integer number 
including zero. This means that photons at these frequencies are generated that
can lead to interference effects. Indeed, inspecting Eq.~(\ref{MQ}), one can realize that the 
two-level emitter with the frequency modulation is reduced to an equivalent system containing 
multiple excited dressed levels, $\omega_{0} \pm |n\omega|$~ $\{n=0,1,2, \cdots\}$, 
decaying to the ground state (see Fig.~\ref{fig1}). When the dressed-state 
splitting is of the order of the cavity mediated radiator's decay rate $\gamma_{0}$ 
then quantum interferences occur among various transition decay paths. For instance,
in Fig.~(\ref{fig1}a) the two decay channels $\omega_{0} \pm \omega$ interfere 
leading to appearance of quantum coherences schematically shown in Fig.~(\ref{fig1}b).
Technically, due to the Bessel function property, $J_{-n}(x)=(-1)^{n}J_{n}(x)$, some of the 
terms from expressions~(\ref{kf}) cancel each other while others add up. To illustrate this 
we chose the simplest case 
$\bar n=1$ and $\delta_{c}=0$, and then the expression for $\gamma(t)$ takes the form:
%%%%%%%%%%%%%%%%%%%%%%%%%%%%%%%%%%%%%%%%%%%%%%%%%%%%%%%%%%%%%%%%%%%%%%%%%%%%%%
\begin{eqnarray}
&{}&\gamma(t)/\gamma_{0} = J^{2}_{0}(\chi) + \kappa^{2}
\bigl\{ J_{1}(\chi)J_{1}(\chi) + J_{-1}(\chi)J_{-1}(\chi) \nonumber \\
&+& 2J_{-1}(\chi)J_{1}(\chi)\cos{[2(\omega t + \phi)]}\bigr\}/(\kappa^{2}+\omega^{2}). 
\label{gmt}
\end{eqnarray}
%%%%%%%%%%%%%%%%%%%%%%%%%%%%%%%%%%%%%%%%%%%%%%%%%%%%%%%%%%%%%%%%%%%%%%%%%%%%%%
The first three terms from Eq.~(\ref{gmt}) describe the spontaneous emission 
processes on the induced transitions: $|e,n=0\rangle \to |g\rangle$ and 
$|e,n= \pm 1\rangle \to |g\rangle$, respectively (see Fig.~\ref{fig1}a). The last term 
in Eq.~(\ref{gmt}) takes into account the cross-correlations among the spontaneously 
decaying channels (see Fig.~\ref{fig1}b, where $n$ denotes a particular sub-level): 
$|e,n=1\rangle \to |g\rangle$ and $|e,n=-1\rangle \to |g\rangle$ or vice versa 
(and, hence, a pre-factor of $2$ there), 
i.e. characterizes quantum decay interference effects \cite{kmek}. On the other hand, 
the cross-decaying correlations among the transition paths: $|e,n=0 \rangle \to |g\rangle$ 
and $|e,n=\pm 1\rangle \to |g\rangle$ reciprocally cancel each other. Obviously, this illustration 
scheme can be extended to $\bar n >1$ (see Fig.~\ref{fig1}b showing the induced coherences 
for $\bar n=2$). These processes together with frequency modulation dressing of the multiple 
decaying rates will lead to a slowing down of the spontaneous emission processes that are also 
absolute phase dependent. Notice that for $\chi \gg 1$ the $n$th- order Bessel function of the 
first kind can be represented as \cite{jan}:
%%%%%%%%%%%%%%%%%%%%%%%%%%%%%%%%%%%%%%%%%%%%%%%%%%%%%%%%%%%%%%%%%%%%%%%%%%%%%%
\begin{eqnarray}
J_{n}(\chi) \approx \sqrt{\frac{2}{\pi \chi}}\cos{(\chi-\pi n/2-\pi/4)},~{\rm when }~ n < \chi.
\label{bsf}
\end{eqnarray}
%%%%%%%%%%%%%%%%%%%%%%%%%%%%%%%%%%%%%%%%%%%%%%%%%%%%%%%%%%%%%%%%%%%%%%%%%%%%%%
The dependence $J_{n}(\chi) \propto 1/\sqrt{\chi}$ will also explain the quenching of the spontaneous decay 
processes for larger modulation depths.  This tendency persists even in the absence of quantum coherences 
due to cross-damping effects. If $n > \chi \gg 1$ the spontaneous decay rate tends to even lower values due 
to the prefactor $1/[\kappa^{2}+ (n\omega)^{2}]$. 
%%%%%%%%%%%%%%%%%%%%%%%%%%%%%%%%%%%%%%%%%%%%%%%%%%%%%%%%%%%%%%%%%%%%%%%%%%%%%%
\begin{figure}[t]
\includegraphics[width = 8cm]{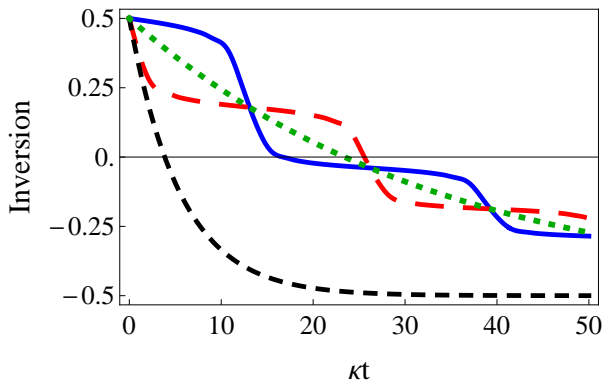}
\caption{\label{fig2} (color online) The time-dependence of the mean-value of the 
inversion operator $\langle S_{z}(t)\rangle$ as a function of $\kappa t$. Here, 
the solid line corresponds to $\chi=50$ and $\phi=0$, the long-dashed one to 
$\chi=50$ and $\phi=\pi/2$ while the dotted curve to $\chi=50$ without 
taking into account quantum coherences. The short-dashed line depicts the 
usual spontaneous decay, i.e. when $\chi=0$. Other parameters are: 
$g=0.3\kappa$, $\omega=0.12\kappa$, $\delta_{c}=0$ and $\kappa=1$.}
\end{figure}
%%%%%%%%%%%%%%%%%%%%%%%%%%%%%%%%%%%%%%%%%%%%%%%%%%%%%%%%%%%%%%%%%%%%%%%%%%%%%%

Figure (\ref{fig2}) shows the population kinetics of an excited two-level emitter 
given by Eq.~(\ref{sz}) for some parameters of interest. Particularly, the 
short-dashed line depicts the typical exponential spontaneous decay low in the 
absence of frequency modulation which is characterized by Eq.~(\ref{sz0}). 
The solid and long-dashed 
curves describe the spontaneous decay processes when the transition frequency is 
modulated with a modulation depth $\chi=50$ and a phase $\phi=0$ or $\phi=\pi/2$, 
respectively. The phase-dependence is a clear evidence of quantum interference effects. 
This occurs for stronger modulation depths $\chi$ and when the frequency of modulation 
$\omega$ is comparable with or less than the single-qubit decay rate $\gamma_{0}$, and 
$\{\omega, \gamma_{0}\} \ll \kappa$. For the sake of comparison, the dotted curve 
characterizes the spontaneous emission behaviour without taking into account the 
quantum coherences due to cross-damping effects (see Fig.~\ref{fig1}a), that is, in Eq.~(\ref{sz0}) we 
have taken 
%%%%%%%%%%%%%%%%%%%%%%%%%%%%%%%%%%%%%%%%%%%%%%%%%%%%%%%%%%%%%%%%%%%%%%%%%%%%%
\begin{eqnarray*}
\tilde \gamma = \sum^{\bar n}_{n=-\bar n}\bar \gamma_{n}J^{2}_{n}(\chi),  ~~~
{\rm ( see ~~Eq.~\ref{kf})}.
\end{eqnarray*}
%%%%%%%%%%%%%%%%%%%%%%%%%%%%%%%%%%%%%%%%%%%%%%%%%%%%%%%%%%%%%%%%%%%%%%%%%%%%%
instead of $\gamma_{0}$. Thus, concluding, the cross-damping effects (i.e. the terms with $n\not=m$ in 
Eq.~\ref{kf} and Eq.~\ref{sz}; see, also, Fig.~\ref{fig1}b) contribute considerably to the final 
spontaneous decay processes (compare solid, dotted and the long-dashed curves in Fig.~\ref{fig2}, respectively). 
%%%%%%%%%%%%%%%%%%%%%%%%%%%%%%%%%%%%%%%%%%%%%%%%%%%%%%%%%%%%%%%%%%%%%%%%%%%%%%
\begin{figure}[t]
\includegraphics[width = 8cm]{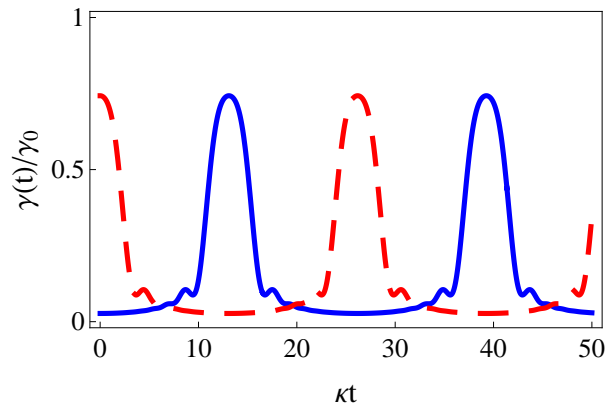}
\caption{\label{fig2g} (color online) The time-dependence of the decay rate $\gamma(t)$
given in Eq.~(\ref{kf}) versus $\kappa t$. Here, the solid line corresponds to $\chi=50$ and 
$\phi=0$ whereas the long-dashed one to $\chi=50$ and $\phi=\pi/2$. 
Other parameters are the same as in Fig.~(\ref{fig2}).}
\end{figure}
%%%%%%%%%%%%%%%%%%%%%%%%%%%%%%%%%%%%%%%%%%%%%%%%%%%%%%%%%%%%%%%%%%%%%%%%%%%%%%

\begin{figure}[t]
\includegraphics[width = 8cm]{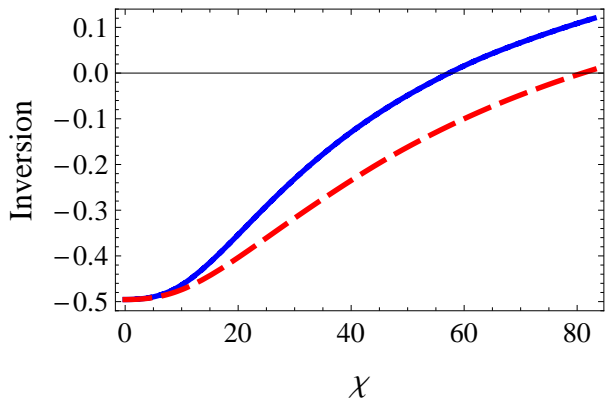}
\caption{\label{fig3} (color online) The mean-value of the inversion operator 
$\langle S_{z}(t)\rangle$ versus the modulation depth $\chi$ when $\kappa t =30$. 
Here, the solid line corresponds to $\phi=0$, while the long-dashed one to 
$\phi=\pi/2$. Other parameters are the same as in Fig.~(\ref{fig2}).}
\end{figure}

The population behaviors shown in Fig.~(\ref{fig2}) are in accordance with the time-dependence form of $\gamma(t)$ 
given in Eq.~(\ref{kf}). The almost decoherence-free plateaus observed in Fig.~(\ref{fig2}) correspond to lower 
values of $\gamma(t)$ (see Fig.~{\ref{fig2g}}). The inhibition of the spontaneous decay can be further improved by increasing the 
modulation depth $\chi$. Therefore, in Fig.~(\ref{fig3}), we have fixed the evolution time at $\kappa t =30$ 
and changed the modulation depth $\chi$ accordingly. At lower modulation amplitudes, or in its absence, the 
qubit is in the ground state at this evolution stage. As it was already mentioned, stronger modulation depths 
contribute to a further slowing of the quantum decoherence. The reason is the interplay between interference 
effects among multiple decay channels described above and the frequency modulation dressing of the corresponding decay rates
(see Eq.~\ref{bsf}). Note however, that the opposite case, i.e. $\omega \gg \kappa$, does 
not show any time- or phase-dependence in the parameters entering in Eq.~(\ref{MQ}) 
or Eq.~(\ref{sz}) and, consequently, no quantum interference effects among the 
different transition pathways occur. This situation was nicely investigated 
in Refs.~\cite{gsa,jano}, respectively.
%%%%%%%%%%%%%%%%%%%%%%%%%%%%%%%%%%%%%%%%%%%%%%%%%%%%%%%%%%%%%%%%%%%%%%%%%%%%%%
%%%%%%%%%%%%%%%%%%%%%%%%%%%%%%%%%%%%%%%%%%%%%%%%%%%%%%%%%%%%%%%%%%%%%%%%%%%%%%

\begin{figure}[t]
\includegraphics[width = 8cm]{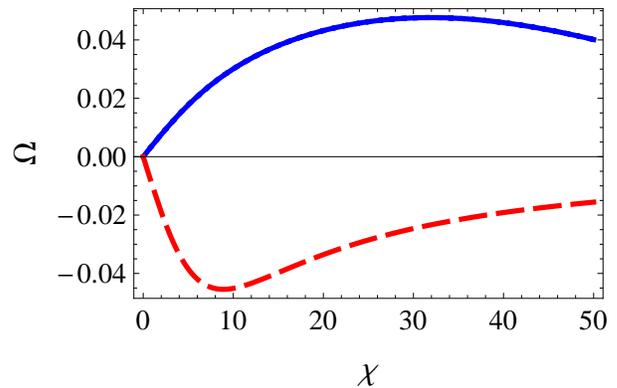}
\caption{\label{fig4} (color online) The frequency shift $\Omega$ (in units of 
$\kappa$) as a function of modulation depth $\chi$ when $\kappa t =10$. Here, 
the solid line corresponds to $\phi=0$, while the long-dashed one to $\phi=\pi/2$. 
Other parameters are the same as in Fig.~(\ref{fig2}).}
\end{figure}

We shall further focus on discussions around the frequency shift due to periodical 
modulation of the transition frequency. Therefore, the frequency shift $\Omega(t)$, 
given in Eq.~(\ref{kf}), is plotted in Fig.~(\ref{fig4}) for particular parameters. 
Here, again, one can observe phase-dependent behaviors due to induced quantum coherences. 
Particularly, when $\bar n=1$ and $\delta_{c}=0$, we have from Eqs.~(\ref{kf},\ref{kdg}):
%%%%%%%%%%%%%%%%%%%%%%%%%%%%%%%%%%%%%%%%%%%%%%%%%%%%%%%%%%%%%%%%%%%%%%%%%%%%%%
\begin{eqnarray}
\Omega(t) = \bar \Omega J_{0}(\chi)\bigl\{J_{1}(\chi)-J_{-1}(\chi)\bigr\}
\cos{(\omega t + \phi)}, \label{shift}
\end{eqnarray}
%%%%%%%%%%%%%%%%%%%%%%%%%%%%%%%%%%%%%%%%%%%%%%%%%%%%%%%%%%%%%%%%%%%%%%%%%%%%%%
where $\bar \Omega = g^{2}\omega/(\kappa^{2}+\omega^{2})$.
One can observe here that the frequency shift is due to cross-correlations 
among the transition channels: $|e,n=0 \rangle \to |g\rangle$ and 
$|e,n=\pm 1\rangle \to |g\rangle$, respectively, i.e. opposite to spontaneous 
emission contributions where these processes cancel out. Evidently, these 
discussions can be generalized for $\bar n >1$. Notice that this frequency 
shift vanishes in the absence of transition frequency modulation at resonance, 
i.e. $\chi=0$, or when $\omega \gg \kappa$ at $\delta_{c}=0$.

For an experimental realization of the proposed scheme we need moderate modulation depths. 
This can be achieved, for instance, in molecular or quantum dot systems possessing a permanent dipole $d_{p}$ 
as it was also mentioned in the article. For $d_{p} \gg d$ and $E_{L}$ being the amplitude 
strength of the applied low-frequency coherent field, one can obtain the necessary modulation depth  
$b \propto (d_{p}\cdot E_{L})$ that is smaller than the transition frequency of the two-level qubit
while $\chi \gg 1$. In asymmetrical quantum dot systems the permanent dipole is proportional to the size 
of the quantum dot and this can be used in engineering of the required model \cite{kibis,okm,shel,pasp_gr}. 
Certain molecules possess this property too, i.e. $d_{p} \gg d$  \cite{kov,kch}.

Finally, while we have considered a broad-band cavity environmental reservoir, the multiple induced decay 
interference approach developed may also be applied for quantized vacuum modes of free-space 
in a related setup \cite{ek}, for instance. There, applying a quantized and sufficiently strong low 
frequency field beyond applicability here, to a two-level atom including far-off resonant states 
in free-space, one can induce quantum interferences among 
few-photon induced transitions. Those process's scaling show an interplay between different 
relevant detunings and applied intensity strengths such that one can stop at a particular 
$\bar n$- photon process \cite{ek} (see Appendix B). 
%%%%%%%%%%%%%%%%%%%%%%%%%%%%%%%%%%%%%%%%%%%%%%%%%%%%%%%%%%%%%%%%%%%%%%%%%%%%%%
%%%%%%%%%%%%%%%%%%%%%%%%%%%%%%%%%%%%%%%%%%%%%%%%%%%%%%%%%%%%%%%%%%%%%%%%%%%%%%

%%%%%%%%%%%%%%%%%%%%%%%%%%%%%%%%%%%%%%%%%%%%%%%%%%%%%%%%%%%%%%%%%%%
\section{Summary}
%%%%%%%%%%%%%%%%%%%%%%%%%%%%%%%%%%%%%%%%%%%%%%%%%%%%%%%%%%%%%%%%%%%
Summarizing, we have demonstrated how quantum decay interference phenomena induced 
among multiple decaying channels occurring due to moderately intense transition frequency 
modulation of a two-level emitter embedded in a broadband electromagnetic field reservoir 
together with frequency modulation dressing of the spontaneous decay rates can suppress 
quantum dissipations due to spontaneous emission. Particularly, phase-dependent low decoherence 
plateaus appear in such a process. Furthermore, a generalized cross-correlated frequency 
shift to the two-level qubit's transition frequency is induced here as well.

%%%%%%%%%%%%%%%%%%%%%%%%%%%%%%%%%%%%%%%%%%%%%%%%%%%%%%%%%%%%%%%%%%%
\acknowledgments
%%%%%%%%%%%%%%%%%%%%%%%%%%%%%%%%%%%%%%%%%%%%%%%%%%%%%%%%%%%%%%%%%%%
We acknowledge the financial support by the German Federal Ministry of Education and 
Research, grant No. 01DK13015, and Academy of Sciences of Moldova, grant No. 13.820.05.07/GF. 
Furthermore, we benefited from useful discussions with Joerg Evers particularly on the 
employed weak-field limit as compared to \cite{ek}. M.M. is grateful to the Theory 
Division of the Max-Planck-Institute for Nuclear Physics from Heidelberg, Germany, 
for kind hospitality.
%%%%%%%%%%%%%%%%%%%%%%%%%%%%%%%%%%%%%%%%%%%%%%%%%%%%%%%%%

%%%%%%%%%%%%%%%%%%%%%%%%%%%%%%%%%%%%%%%%%%%%%%%%%%%%%%%%%
\appendix
%%%%%%%%%%%%%%%%%%%%%%%%%%%%%%%%%%%%%%%%%%%%%%%%%%%%%%%%%
\section{Master equation with a modulated transition in free space}
%%%%%%%%%%%%%%%%%%%%%%%%%%%%%%%%%%%%%%%%%%%%%%%%%%%%%%%%%
In this Appendix, we shall focus on the spontaneous decay of an excited two-level emitter with modulated transition frequency via 
usual vacuum modes of the EMF reservoir. The purpose is to show that an unshaped vacuum is not 
sufficient for the spontaneous emission suppression investigated here. For convenience 
we derive the modulation Hamiltonian entering in Eq.~(\ref{Hh}). For this, one considers that a moderately 
strong low-frequency coherent field is applied to a two-state atom being initially in its excited state. 
The Hamiltonian of this process is:
%%%%%%%%%%%%%%%%%%%%%%%%%%%%%%%%%%%%%%%%%%%%%%%%%%%%%%%%%%%%%%%%%%%
\begin{eqnarray}
H &=&\sum_{k}\hbar\omega_{k}a^{\dagger}_{k}a_{k}+ \hbar \omega_{0}S_{z} + \hbar\Omega\cos{(\omega t)}\bigl(S^{+} + S^{-}\bigr) \nonumber \\
&+&i\sum_{k}(\vec g_{k}\cdot \vec d)\bigl(a^{\dagger}_{k}-a_{k}\bigr)\bigl(S^{+} + S^{-}\bigr).
\label{Hp}
\end{eqnarray}
%%%%%%%%%%%%%%%%%%%%%%%%%%%%%%%%%%%%%%%%%%%%%%%%%%%%%%%%%%%%%%%%%%%
Here, $\omega_{0}$ is the transition frequency among the states $|e\rangle \leftrightarrow |g\rangle$ while $\Omega$ is the corresponding Rabi frequency
with $\omega$ being the frequency of the external applied low-frequency coherent field. The atom-vacuum coupling strength is: 
$\vec g_{k}=\sqrt{2\pi\hbar\omega_{k}/V}\vec e_{\lambda}$, where $V$ is quantized volume while $\vec e_{\lambda}$ is the polarization vector with 
$\{\lambda =1,2\}$.

We apply a unitary transformation to the Hamiltonian (\ref{Hp}), i.e. $H \equiv U(t)HU^{-1}(t)$, where 
$U(t)=\exp{\bigl[i\frac{\Omega}{\omega}\sin{(\omega t)}\bigl(S^{+} + S^{-}\bigr)\bigr]}$. This transformation is useful as it allows to represent the 
Hamiltonian via $n-$photon processes involved, namely:
%%%%%%%%%%%%%%%%%%%%%%%%%%%%%%%%%%%%%%%%%%%%%%%%%%%%%%%%%%%%%%%%%%%
\begin{eqnarray}
H &=& \hbar \omega_{0}S_{z}\sum^{\infty}_{n=0}J_{n}(\rho)\cos{(n\omega t)}\bigl(1+(-1)^{n}(1-\delta_{n,0})\bigr) \nonumber \\
&+&\frac{i}{2}\hbar\omega_{0}(S^{-}-S^{+})\sum^{\infty}_{n=1}J_{n}(\rho)\sin{(n\omega t)}\bigl(1 - (-1)^{n}\bigr) \nonumber \\
&+&\sum_{k}\hbar\omega_{k}a^{\dagger}_{k}a_{k} + i\sum_{k}(\vec g_{k}\cdot \vec d)\bigl(a^{\dagger}_{k}-a_{k}\bigr)\bigl(S^{+} + S^{-}\bigr).
\nonumber \\
\label{Hpf}
\end{eqnarray}
%%%%%%%%%%%%%%%%%%%%%%%%%%%%%%%%%%%%%%%%%%%%%%%%%%%%%%%%%%%%%%%%%%%
where $\rho=2\Omega/\omega$. It is easy to observe that the even-photon processes with $n=2,4,\cdots$ correspond to modulation of the 
transition frequency, while the odd-photon processes, i.e. $n=1,3,\cdots$  lead to induced transitions among the involved energy levels. For our purposes 
one requires $\rho \ll 1$ as well as $\omega \ll \omega_{0}$. Under these restrictions, i.e. taking into account that: 
%%%%%%%%%%%%%%%%%%%%%%%%%%%%%%%%%%%%%%%%%%%%%%%%%%%%%%%%%%%%%%%%%%%
\begin{eqnarray*}
J_{n}(\rho) \approx \rho^{n}\biggl\{\frac{2^{-n}}{\Gamma(1+n)} - \frac{2^{-2-n}\rho^{2}}{(1+n)\Gamma(1+n)} + O[\rho]^{4}\biggr\},
\end{eqnarray*}
%%%%%%%%%%%%%%%%%%%%%%%%%%%%%%%%%%%%%%%%%%%%%%%%%%%%%%%%%%%%%%%%%%%
the working Hamiltonian is:
%%%%%%%%%%%%%%%%%%%%%%%%%%%%%%%%%%%%%%%%%%%%%%%%%%%%%%%%%%%%%%%%%%%
\begin{eqnarray}
H &=& \sum_{k}\hbar\omega_{k}a^{\dagger}_{k}a_{k}+ \hbar \bigl(\omega_{0} + b\cos{(2\omega t)} + b'\cos{(4\omega t)}\bigr) S_{z} \nonumber \\
&+& i\sum_{k}(\vec g_{k}\cdot \vec d)\bigl(a^{\dagger}_{k}-a_{k}\bigr)\bigl(S^{+} + S^{-}\bigr).
\label{Hpt}
\end{eqnarray}
%%%%%%%%%%%%%%%%%%%%%%%%%%%%%%%%%%%%%%%%%%%%%%%%%%%%%%%%%%%%%%%%%%%
Here, $b = \omega_{0}\rho^{2}/4$, $b' = \omega_{0}\rho^{4}/192$ and $\omega_{0} \equiv \omega_{0}(1-\rho^{2}/4)$. Thus, we have 
considered that the frequency modulation takes place via two- and four-photon processes, simultaneously, while higher-photon effects are negligible because 
one can always select a system with $\omega_{0}\rho^{n}/\omega \ll \omega_{0}\rho^{n-2}/\omega$, and for an even $n$ with $n>4$. Furthermore, the 
induced transitions through odd-photon processes among the involved energy levels do not occur because of the off-resonance and, therefore, are ignored 
here (i.e. the second line of the Hamiltonian \ref{Hpf}).
%%%%%%%%%%%%%%%%%%%%%%%%%%%%%%%%%%%%%%%%%%%%%%%%%%%%%%%%%%%%%%%%%%%%%%%%%%%%%%
\begin{figure}[t]
\centering
\includegraphics[width =8.8cm]{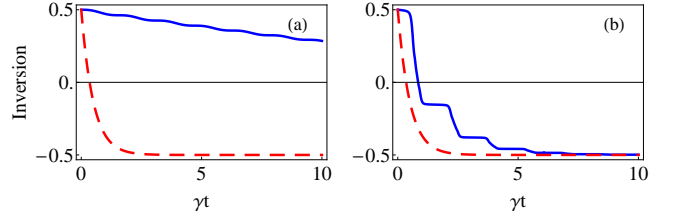}
\caption{\label{fig5} (color online) The free-space mean-value of the inversion operator 
$\langle S_{z}(t)\rangle$ versus $\gamma t$. Here, $\rho=0.2$, $\omega/\gamma=1$, $\phi=0$ while $\omega_{0}/\omega=2\cdot 10^{4}$ and, 
hence, $\chi=100$. (a) $n_{0}=1$; (b) $n_{0}=50$. The dashed line shows the usual free-space spontaneous decay low in the absence of frequency 
modulation.}
\end{figure}
%%%%%%%%%%%%%%%%%%%%%%%%%%%%%%%%%%%%%%%%%%%%%%%%%%%%%%%%%%%%%%%%%%%%%%%%%%%%%%

The master equation describing the spontaneous decay of a two-level radiator with modulated transition frequency in free-space according to the Hamiltonian 
(\ref{Hpt}) and in the Born-Markov, dipole and rotating wave approximations has the form:
%%%%%%%%%%%%%%%%%%%%%%%%%%%%%%%%%%%%%%%%%%%%%%%%%%%%%%%%%%%%%%%%%%%
\begin{eqnarray}
\frac{d}{dt}\langle Q(t)\rangle = -\bigl(\gamma_{f}(t) - i\Omega_{f}(t)\bigr)\langle S^{+}[S^{-},Q]\rangle + H.c., \nonumber \\
\label{MQQ}
\end{eqnarray}
%%%%%%%%%%%%%%%%%%%%%%%%%%%%%%%%%%%%%%%%%%%%%%%%%%%%%%%%%%%%%%%%%%%
where
%%%%%%%%%%%%%%%%%%%%%%%%%%%%%%%%%%%%%%%%%%%%%%%%%%%%%%%%%%%%%%%%%%%
\begin{eqnarray*}
\gamma_{f}(t) - i\Omega_{f}(t) = \sum^{n_{0}}_{n,n'=-n_{0}}\sum^{m_{0}}_{m,m'=-m_{0}}J_{n}(\chi)J_{n'}(\chi)J_{m}(\chi') \nonumber \\
\times J_{m'}(\chi')\bigl(\gamma_{0n'm'} - i\Omega_{0n'm'} \bigr)e^{-2i\omega t(n-n')}e^{-4i\omega t(m-m')},
\end{eqnarray*}
%%%%%%%%%%%%%%%%%%%%%%%%%%%%%%%%%%%%%%%%%%%%%%%%%%%%%%%%%%%%%%%%%%%
with
%%%%%%%%%%%%%%%%%%%%%%%%%%%%%%%%%%%%%%%%%%%%%%%%%%%%%%%%%%%%%%%%%%%
\begin{eqnarray*}
\Omega_{0n'm'} &=& \sum_{k}\frac{(\vec g_{k}\cdot \vec d)^{2}}{\hbar^{2}}P\frac{1}{\omega_{k} - \omega_{0} - 2n'\omega -4m'\omega},  \nonumber \\
\gamma_{0n'm'} &=& \gamma(1 + \frac{2n'\omega}{\omega_{0}} + \frac{4m'\omega}{\omega_{0}})^{3}, \nonumber \\
{\rm and} &{}& \chi=b/(2\omega), ~~{\rm while}~~ \chi'=b'/(4\omega).
\end{eqnarray*}
%%%%%%%%%%%%%%%%%%%%%%%%%%%%%%%%%%%%%%%%%%%%%%%%%%%%%%%%%%%%%%%%%%%
Here, $P$ is the Cauchy principal value, while $2\gamma = 4d^{2}\omega^{3}_{0}/(3\hbar c^{3})$ is the free-space single-atom spontaneous decay rate in the 
absence of frequency modulation \cite{kmek}. 

In the following, we consider that $\chi \gg 1$ while $\chi' \ll 1$, i.e. the frequency modulation via simultaneous four-photon processes are negligible. 
This can be achieved in the weak field regime, for example, when $\rho=2\cdot 10^{-1}$ and $\omega_{0}/\omega=2\cdot 10^{4}$, that is, one has $\chi=10^{2}$ while 
$\chi' \approx 4.2\cdot 10^{-2}$. For  this reason we do not expect to obtain the quantum interference effects without a cavity like in \cite{ek}. Thus, the motivation to keep the modulation of the transition frequency due to a simultaneous four-photon process in the 
Hamiltonian (\ref{Hpt}) as well as in the master equation (\ref{MQQ}), i.e. the term proportional to $b'$, was to show that spontaneous decay via the 
induced absorption/emission of four photons is taking place also due to transition frequency modulation via a two-photon process described by $b$ 
(for instance when $n_{0}=2$). This latter process is more probable than the corresponding one due to $b'$. 

Figure (\ref{fig5}) shows the spontaneous decay law of an excited atom in free space, i.e. $\langle S_{z}(t)\rangle=\exp{[-2\Gamma_{f}(t)]}-1/2$ 
where $\Gamma_{f}(t)=\int^{t}_{0}\gamma_{f}(\tau)d\tau$ with $m=m'=0$,  when $n_{0}=1$ and $n_{0}=50$, respectively, and for some particular 
parameters of interest. While the spontaneous decay is clearly slowed down for $n_{0} \ll \chi >1$ one can not predict what is the particular value of $n_{0}$ 
which would be realized in a real experiment. Furthermore, larger values of $n_{0}$ with a fixed $\chi=100$ do not lead to spontaneous emission inhibition. 
Thus, it is unlikely that the spontaneous emission will be inhibited in free-space via transition frequency modulation of an excited two-level emitter unless and 
until we are not able to select a particular $n_{0}$-photon process.

%%%%%%%%%%%%%%%%%%%%%%%%%%%%%%%%%%%%%%%%%%%%%%%%%%%%%%%%%
%%%%%%%%%%%%%%%%%%%%%%%%%%%%%%%%%%%%%%%%%%%%%%%%%%%%%%%%%
\section{The spontaneous decay modification via applying a 
strong quantized low frequency field}
%%%%%%%%%%%%%%%%%%%%%%%%%%%%%%%%%%%%%%%%%%%%%%%%%%%%%%%%%
In Ref.~\cite{ek} a related mechanism was discussed where an 
effective two level system was driven with a quantum field 
with frequency $\omega$ of the order of the linewidth of the 
considered transition.  Since in reality the low-frequency field 
will also couple off-resonantly to all transitions in the atom, 
one can justify a Hamilton operator with extra interference terms as 
indicated in detail in the Ref.~\cite{ek} with the consequence of 
possible spontaneous emission elimination. The physics behind atom 
pumping with a classical low frequency field  in the weak field 
approximation or a quantized low frequency field consists in the fact 
that in the first case only modulation of the transition 
frequency occurs which does not affect the spontaneous emission in free 
space, while in the second case additional transitions are induced that 
may interfere leading to modification of the spontaneous decay.
Particularly, in \cite{ek} a quantized low frequency strong field was 
interacting with a two-level atom being initially in its excited state 
$|2\rangle$. In order to adequately describe the system the far-off resonant 
coupling to the other states was included as well. 

In what follows, we recall in a somewhat different simplified way how the additional 
terms due to off-resonant coupling may appear in the interaction Hamiltonian. 
If as an example one considers a 
two-level atom with one extra far off-resonant state $|3\rangle$ and with its energy 
levels from down to up denoted as $|1\rangle, |2\rangle$ 
and $|3\rangle$, respectively, then the quantized low frequency applied field 
will induce additional transitions via the path $|2\rangle \to |3\rangle \to 
|1\rangle$. In order to involve the corresponding terms one need to go beyond 
the weak low-frequency field approximation applied in the main part of this work.
The interaction Hamiltonian responsible for these terms is:
%%%%%%%%%%%%%%%%%%%%%%%%%%%%%%%%%%%%%%%%%%%%%%%%%%%%%%%%%
\begin{eqnarray}
H_{I} \approx -2i(c + c^{\dagger})\sum_{k}
(\vec g_{k}\cdot \vec d_{31})\alpha_{32}
(a^{\dagger}_{k}S^{-} - a_{k}S^{+}). \label{b1}
\end{eqnarray}
%%%%%%%%%%%%%%%%%%%%%%%%%%%%%%%%%%%%%%%%%%%%%%%%%%%%%%%%%
Here $\alpha_{ij} \propto d_{ij}E_{L}$ are the corresponding coefficients 
due to strong external pumping and obtained after elimination of the 
far-laying excited level $|3\rangle$, while $c^{\dagger}(c)$ is the photon 
creation (annihilation) 
operator for the external low frequency quantized intense field. 
One can show (see the last entry of \cite{ek}) that the Hamiltonian (\ref{b1})
does not affect the transition $|2\rangle \to |1\rangle$ when it is dipole
allowed. Therefore, the spontaneous emission modification scheme described 
in this part applies to dipole-forbidden atomic transitions. Since 
in a three-level atomic system one transition should be dipole-forbidden 
it results that the decay on $|2\rangle \to |1\rangle$ atomic transition 
may indeed be dipole-forbidden. 
The excited state population will be described by the following 
equation:
%%%%%%%%%%%%%%%%%%%%%%%%%%%%%%%%%%%%%%%%%%%%%%%%%%%%%%%%%
\begin{eqnarray}
\frac{d}{dt}\langle S_{22}\rangle = - \gamma_{b}\langle S_{22}\rangle 
- \gamma_{a} \langle S_{22}(c e^{-i\omega t} + c^{\dagger}e^{i\omega t})^{2}\rangle.
\end{eqnarray}
%%%%%%%%%%%%%%%%%%%%%%%%%%%%%%%%%%%%%%%%%%%%%%%%%%%%%%%%%
Here, $\gamma_{b}$ is the two-photon spontaneous decay rate on transition 
$|2\rangle \to |1\rangle$, while $\gamma_{a}$ is the corresponding decay 
involving additional upper states. One can observe that the strong quantized 
field indeed modifies the upper state population. Notice that a strong classical 
low frequency laser field will only modulate the frequency on a dipole-allowed 
transition and, correspondingly, the spontaneous emission will not be modified 
in free space (see also Appendix A). 
For dipole-forbidden $|1\rangle \leftrightarrow |2\rangle$ transitions the off-resonant 
coupling to another state $|3\rangle$ will also modulate the transition frequency as 
well as induce transitions to the ground state $|1\rangle$ based on a Hamiltonian of 
the next form:
%%%%%%%%%%%%%%%%%%%%%%%%%%%%%%%%%%%%%%%%%%%%%%%%%%%%%%%%%
\begin{eqnarray}
H_{I} \approx -i\sum_{k}(\vec g_{k}\cdot \vec d_{31})\alpha_{32}
(a^{\dagger}_{k}S^{-} - 
a_{k}S^{+})\cos(\omega t).
\end{eqnarray}
%%%%%%%%%%%%%%%%%%%%%%%%%%%%%%%%%%%%%%%%%%%%%%%%%%%%%%%%%
The spontaneous decay rate on the $|2\rangle \to |1\rangle$ atomic transition 
may be $\gamma_{b} + \alpha^{2}_{32}\gamma_{31}$. Therefore, the classical 
off-resonant coupling to another state may not change significantly the decay rate 
because $\alpha_{32} < 1$.

In a two-level atomic system with two extra far off-resonant states, the 
spontaneous emission can be modified due to an applied strong quantized 
low frequency field even on a dipole-allowed
$|2\rangle \leftrightarrow |1\rangle$ transition in accordance with the results 
given in Ref.~\cite{ek}. For instance, one of the terms in the total Hamiltonian 
describing the decay via the path $|2\rangle \to |4\rangle \to |3\rangle \to |1\rangle$ 
after the elimination of the additional higher energy levels is:
%%%%%%%%%%%%%%%%%%%%%%%%%%%%%%%%%%%%%%%%%%%%%%%%%%%%%%%%%%%%%%%%%%%%%%%%%%%%%%%
\begin{eqnarray}
H_{I} \approx i(c + c^{\dagger})^{2} \sum_{k}(\vec g_{k}\cdot \vec d_{31})\alpha_{24}\alpha_{43}
(a^{\dagger}_{k}S^{-} - a_{k}S^{+}). \nonumber \\
\end{eqnarray}
%%%%%%%%%%%%%%%%%%%%%%%%%%%%%%%%%%%%%%%%%%%%%%%%%%%%%%%%%%%%%%%%%%%%%%%%%%%%%%%%
Such terms are responsible for the spontaneous decay modification in free-space 
through a quantized and strong low frequency applied field. On the other side, 
in an idealised situation of ignoring the presence of any other far-off resonant 
states in the atomic system and when neglecting strong field terms of the low 
frequency field, the presence of such interference terms could not be 
reproduced via a classical pumping field. 
%%%%%%%%%%%%%%%%%%%%%%%%%%%%%%%%%%%%%%%%%%%%%%%%%%%%%%%%%
%%%%%%%%%%%%%%%%%%%%%%%%%%%%%%%%%%%%%%%%%%%%%%%%%%%%%%%%%%%%%%%%%%%

%%%%%%%%%%%%%%%%%%%%%%%%%%%%%%%%%%%%%%%%%%%%%%%%%%%%%%%%%%%%%%%%%%%%%%%%%%%%%%%%%%%%%%%%%%%
\end{document}
%%%%%%%%%%%%%%%%%%%%%%%%%%%%%%%%%%%%%%%%%%%%%%%%%%%%%%%%%%%%%%%%%%%%%%%%%%%%%%%%%%%%%%%%%%%%